\documentstyle[prl,aps,twocolumn,epsfig]{revtex}
\def\be{\begin{equation}}
\def\ee{\end{equation}}
\def\bea{\begin{eqnarray}}
\def\eea{\end{eqnarray}}

\begin{document}

\title{Asymmetric nuclear matter: the role of
the isovector scalar channel}

\author{B. Liu$^{1,2}$, V. Greco$^{1}$, V. Baran$^{1,3}$, M. Colonna$^1$
and M. Di Toro$^1$}

\address{$^1$ Laboratorio Nazionale del Sud, Via S. Sofia 44,
I-95123 Catania, Italy\\ and University of Catania}
\address{$^{2}$ Institute of High Energy Physics, Chinese Academy of Sciences,
Beijing 100039, China}
\address{$^{3}$ NIPNE-HH, Bucharest, Romania}

\maketitle

\begin{abstract}
We try to single out some qualitative new effects of the coupling to the
$\delta$-isovector-scalar meson introduced in a minimal way
in a phenomenological hadronic field theory. Results for the equation
of state ($EOS$) and the phase diagram of asymmetric nuclear matter
($ANM$) are discussed. We stress the consistency of the $\delta$-coupling
introduction in a relativistic approach.

New contributions
to the slope and curvature of the symmetry energy and the neutron-proton
effective mass splitting appear particularly interesting.

A more repulsive $EOS$ for neutron matter at high baryon densities is
expected.

Effects on new critical properties of warm $ANM$, mixing of mechanical 
and chemical instabilities and isospin distillation, are also presented.
The $\delta$ influence is mostly on the {\it isovectorlike}
collective response.

The results are largely analytical and this makes the physical meaning
quite transparent. Implications for nuclear structure properties of drip-line
nuclei and for reaction dynamics with Radioactive Beams are finally
pointed out.

\end{abstract}

\vspace{1cm}

%{\bf PACS numbers :}$~~$21.65.+f, 21.30.Fe, 25.70.Pq, 25.75.-q\\

\hspace{-\parindent}PACS numbers: 21.65.+f, 21.30.Fe, 25.70.Pq, 24.10.Jv

\section{Introduction}

Hadronic effective field theories [quantum hadrodynamics, $QHD$]
represent a significant improvement in the understanding of static
and dynamical properties of nuclear matter and finite nuclei, described
as strongly interacting systems of baryons and mesons. The main application
of this approach has been the Relativistic Mean Field theory ($RMF$)
\cite{wal74,sewa86}, extremely successful in nuclear structure studies
\cite{suto94,lari97,sha00}. 

In recent years, due to the new possibilities opened by the Radioactive 
Beam facilities, the interest has moved towards a microscopic
description of asymmetric nuclear systems, from the stability
of drip-line nuclei to the phase diagram of asymmetric nuclear matter.
The connection to astrophysics problems, supernovae explosions and 
neutron stars, is also quite evident \cite{sumi94,le96,pra97,bom01}.
The isovector channel has been introduced through a coupling to
the charged vector $\rho$-meson. In the Hartree approximation
of the $RMF$ approach it leads to a simple linear increase of
the symmetry term with the baryon density, without neutron/proton
effective mass splitting \cite{yst98}. A complete study of the
corresponding phase diagram for asymmetric nuclear matter has
been performed in ref. \cite{mu95}, with a thorough discussion of the
new qualitative features of the liquid-gas phase transition.

In this work we focus our attention on the introduction 
of the coupling to the isovector scalar channel, through the
exchange of a virtual charged $\delta [a_0(980)]$ meson. In this way,
as well know in the isoscalar channel, we recover the genuine
structure of the relativistic interactions, where one has the
balance between scalar (attractive) and vector (repulsive)
"potentials". Actually the $\delta$-meson exchange is an essential 
ingredient of all $NN$ realistic potentials and its inclusion
in the $QHD$ scheme has been already suggested \cite{ku97}, also
on the basis of a relativistic Brueckner theory \cite{sst97,jole98,hole01}.
Some first results of structure calculations for exotic nuclei are
showing the importance of the $\delta$-dynamics for the stability
conditions of drip-line nuclei \cite{hole01,legm01}.

Aim of this paper is to present, within a $RMF$ approach, the main 
expected effects of the $\delta$-field on symmetry properties
of the nuclear system, from $EOS$ to $n,p$-mass splitting, and
in particular on the nuclear response in unstable regions. We remind that the 
study of symmetric and asymmetric nuclear matter
properties under extreme conditions is of great relevance
for the understanding of the nuclear interaction in the medium.
In particular a liquid-gas phase transition may occur in warm and 
dilute matter 
produced in heavy-ion collisions (multifragmentation events). 
This is one of the interesting 
open problems in theoretical and experimental nuclear physics.
Isospin effects on heavy ion collisions (fragmentation and collective flows)
have been largely discussed in recent years, see the reviews in
\cite{nn00,nova01}. In this work we will try to offer a unified
picture of the $\delta$-meson exchange influence on nuclear structure
and dynamics.

In order to simplify the analysis and to pin down the most
direct effects of the $\delta$ contributions we will follow
a quite reduced version of the $RMF$ approach, yet including all
the isoscalar $(\sigma,\omega)$ and isovector $(\delta,\rho)$
fields. 
Non-linear self-interaction terms are introduced only in the isoscalar
scalar $\sigma$-channel, essential in order to get right incompressibility
parameter at normal density $\rho_0$ \cite{bobo77,bd91}. 
Such approximation scheme of minimal self-interacting terms is
actually physically well justified in the baryon density
range of interest here, roughly up to $2.5\rho_0$ \cite{self}.
Moreover it is consistent with the $RMF$ assumptions of neglecting
retardation and finite range effects in the field dynamics.

We like to stress the presence of several points of interest on the isovector
channel results in this baryon density region \cite{pera95}:
\begin{itemize}
\item{
The slope of the symmetry term, the {\it Symmetry Pressure}, just
below $\rho_0$ is of relevant importance for nuclear structure,
being directly linked to the thickness of the neutron skin in
n-rich (stable and/or unstable) nuclei, see the recent discussions
in \cite{bro00,typ01,hor01}. Jointly to the $n-p$ effective masses
this information appears  essential for the stability of
drip-line nuclei \cite{hole01,legm01}.

Dissipative reaction mechanisms
with asymmetric ions seem to be also quite sensitive to the same quantity
 \cite{fab98,eri98}.}
\item{ The nature of chemical instabilities and in general critical
properties of warm and dilute asymmetric nuclear matter 
\cite{mu95,bao197,cdl98,bar98,bar01}.}
\item{ Isospin effects on the dynamics of heavy ion collisions
at intermediate energies, see \cite{nova01}. 
Reaction studies represent a  very sensitive tool
to test transport properties of symmetry effects, $n,p$ chemical potentials
and effective masses \cite{bao00,ditcr,ditba}.}
\end{itemize}

In this paper, we first determine the model parameters by 
fitting the properties of
the symmetric and asymmetric nuclear matter at $T=0$.
We then extend the investigation 
to finite temperature. In particular we derive the new
boundary regions for mechanical and chemical instabilities 
and their dependence on the various mesons entering in the theory.

This paper is organized as follows.
In Sec. 2, the equation of state for nuclear matter at finite temperature 
is derived. Symmetry energies and effective masses are discussed in
Sec.3, where the neutron matter $EOS$ is also evaluated. 
The mechanical and chemical instabilities are 
studied in Sec. 4. General comments
as well as a summary of the main conclusions 
are presented in the last Section.

\section{Equation of state for nuclear matter at finite temperature}
The starting point is the relativistic Lagrangian density
of an interacting many-particle system consisting of nucleons,
isoscalar (scalar $\sigma$, vector $\omega$) and isovector (scalar 
$\delta$, vector $\rho$) mesons:

\begin{eqnarray}\label{eq.1}
{\cal L } &=& \bar{\psi}
[i\gamma_{\mu}\partial^{\mu}-(M_{N}- g_{\sigma}\phi
-g_{\delta}\vec{\tau}\cdot\vec{\delta}) \nonumber \\
&& -g{_\omega}\gamma_\mu\omega^{\mu}-g_\rho\gamma^{\mu}\vec\tau\cdot
\vec{b_\mu}]\psi \nonumber \\
&& +\frac{1}{2}(\partial_{\mu}\phi\partial^{\mu}\phi-m_{\sigma}^2\phi^2)
-U(\phi)+\frac{1}{2}m^2_{\omega}\omega_{\mu} \omega^{\mu} \nonumber \\
&& +\frac{1}{2}m^2_{\rho}\vec{b_{\mu}}\cdot\vec{b^{\mu}} 
+\frac{1}{2}(\partial_{\mu}\vec{\delta}\cdot\partial^{\mu}\vec{\delta}
-m_{\delta}^2\vec{\delta^2}) \nonumber \\
&& -\frac{1}{4}F_{\mu\nu}F^{\mu\nu}
-\frac{1}{4}\vec{G}_{\mu\nu}\vec{G}^{\mu\nu},
\end{eqnarray}

 Minimal self-interacting terms,
as discussed in the introduction, are included only in the $\sigma$-channel.

Here $\phi$ is the $\sigma$-meson field,
$\omega_{\mu}$ the $\omega$-meson field, $\vec{b_{\mu}}$ the
charged $\rho$ meson field, $\vec{\delta}$ the isovector scalar field 
of the $\delta$-meson.
We define
$F_{\mu\nu}\equiv\partial_{\mu}\omega_{\nu}-\partial_{\nu}\omega_{\mu}$,
$\vec{G}_{\mu\nu}\equiv\partial_{\mu}\vec{b_{\nu}}-\partial_{\nu}\vec{b_{\mu}}$.
$U(\phi)$ is the nonlinear potential of the $\sigma$ meson :
$U(\phi)=\frac{1}{3}a\phi^{3}+\frac{1}{4}b\phi^{4}$.

The field equations in mean field approximation ($RMF$) are:

\begin{eqnarray}\label{eq.2}
(i\gamma_{\mu}\partial^{\mu}-(M_{N}- g_{\sigma}\phi
-g_\delta{\tau_3}{\delta_3})
-g_\omega\gamma^{0}{\omega_0} \nonumber \\
 -g_\rho\gamma^{0}{\tau_3}{b_0})\psi=0 \nonumber \\
m_{\sigma}^2\phi+ a{{\phi}^2}+ b{{\phi}^3}=\bar\psi\psi
=g_\sigma{\rho}_S \nonumber \\
m^2_{\omega}\omega_{0}=g_\omega\bar\psi{\gamma^0}\psi=g_\omega{\rho}_B 
 \nonumber \\
m^2_{\rho}b_{0}=g_\rho\bar\psi{\gamma^0}\tau_3\psi=g_\rho\rho_{B3} 
 \nonumber \\
m^2_{\delta}\delta_3=g_{\delta}\bar\psi\tau_3\psi=g_{\delta}\rho_{S3},
\end{eqnarray}
where $\rho_{B3}=\rho_{Bp}-\rho_{Bn}$ and 
$\rho_{S3}=\rho_{Sp}-\rho_{Sn}$, $\rho_B$ and $\rho_S$ being
the baryon and the scalar densities, respectively. 

Neglecting the derivatives of mesons fields, in the mean field
approximation the energy-momentum tensor 
is given by

\begin{eqnarray}\label{eq.3}
T_{\mu\nu}=i\bar{\psi}\gamma_{\mu}\partial_{\nu}\psi+[\frac{1}{2} 
 m_{\sigma}^2\phi^2+U(\phi)+\frac{1}{2}m_{\delta}^2\vec{\delta^2}
 \nonumber \\
-\frac{1}{2}m^2_{\omega}\omega_{\lambda} \omega^{\lambda}
-\frac{1}{2}m^2_{\rho}\vec{b_{\lambda}}\vec{b^{\lambda}}]g_{\mu\nu}.
\end{eqnarray}

The properties of nuclear matter at finite temperature are described by
the thermodynamic potential $\Omega$.
>From statistical mechanics for a system in a volume V we define
$\Omega= -pV =-{1\over{\beta}} lnZ$ \cite{sewa86},
where $\beta$ the inverse of temperature, $\beta$=1/T,
and $Z$ is the grand partition function given by
$Z=Tr[e^{-\beta(\hat{H}-\Sigma_i(\mu_i\hat{B}_i))}]$.  
Here $\hat{H}$ is the hamiltonian operator, $\hat{B}_i$ and $\mu_i$
are respectively nucleon number
operators and thermodynamic chemical potentials, $(i=p,n)$.
The equation of state for nuclear matter at finite temperature
can be obtained from the thermodynamic potential $\Omega$.

The energy density has the form

\begin{eqnarray}\label{eq.4}
\epsilon=
\sum_{i=n,p}{2}\int \frac{{\rm d}^3k}{(2\pi)^3}E_{i}^\star(k)
(n_i(k)+\bar{n_i}(k))
+\frac{1}{2}m_\sigma^2\phi^2
 \nonumber \\
+U(\phi)+\frac{1}{2}m_\omega^2\omega_0^2
+\frac{1}{2}m_{\rho}^2 b_0^2
+ \frac{1}{2}m_{\delta}^2\delta_3^2,
\end{eqnarray}

and the pressure is

\begin{eqnarray}\label{eq.5}
P =\sum_{i=n,p} \frac{2}{3}\int \frac{{\rm d}^3k}{(2\pi)^3}
\frac{k^2}{E_{i}^\star(k)}
(n_i(k)+\bar{n_i}(k)) -\frac{1}{2}m_\sigma^2\phi^2 \nonumber \\
- U(\phi)+
\frac{1}{2}m_\omega^2\omega_0^2
+\frac{1}{2}m_{\rho}^2{b_0^2}
-\frac{1}{2}m_{\delta}^2\delta_3^2,
\end{eqnarray}
where
${E_i}^\star=\sqrt{k^2+{{M_i}^\star}^2}$. The nucleon effective masses
are defined as 
\begin{equation}\label{eq.6}
{M_i}^\star=M_{N}-g_\sigma\phi\mp g_\delta\delta_3~~~ (-~proton, +~neutron).
\end{equation}
The $n_i(k)$ and $\bar{n_i}(k)$ in Eqs.
(4)-(5) are the fermion and antifermion distribution functions for protons
($i=p$) and neutrons ($i=n$):

\begin{eqnarray}\label{eq.7}
n_i(k)=\frac{1}{1+\exp\{({E_i}^\star(k)-{\mu_i^\star})/T \} }\,,
\end{eqnarray}
and
\begin{eqnarray}\label{eq.8}
\bar{n_i}(k)=\frac{1}{1+\exp\{({E_i}^\star(k)+{\mu_i^\star})/T \} }.
\end{eqnarray}

The effective chemical potentials $\mu_i^\star$  are given in terms
of the vector meson mean fields 

\begin{eqnarray}\label{eq.9}
{\mu_i}=\mu_i^\star  - g_\omega\omega_0\mp g_{\rho}b_0~~~(-~proton, +~neutron),
\end{eqnarray}

where $\mu_i$ are the thermodynamical chemical potentials 
$\mu_i=\partial\epsilon/\partial\rho_i$. At zero temperature they
reduce to the Fermi energies $E_{Fi} \equiv \sqrt{k_{Fi}^2+{M_i^\star}^2}$.

The same results can be directly obtained from the expectation value of
the energy-momentum tensor, showing the thermodynamical consistency of
the mean field approximation \cite{sewa86}.

By using the field equations for mesons, the equations of state for thermal
matter can be rewritten as

\begin{eqnarray}\label{eq.10}
\epsilon=
\sum_{i=n,p}{2}\int \frac{{\rm d}^3k}{(2\pi)^3}E_{i}^\star(k)
(n_i(k)+\bar{n_i}(k))
+\frac{1}{2}m_\sigma^2\phi^2 \nonumber \\
+U(\phi)
+\frac{g_{\omega}^2}{2m_\omega^2}\rho_B^2
+\frac{g_{\rho}^2}{2m_{\rho}^2}\rho_{B3}^2
+ \frac{g_{\delta}^2}{2m_{\delta}^2}\rho_{S3}^2,
\end{eqnarray}

and

\begin{eqnarray}\label{eq.11}
P =\sum_{i=n,p} \frac{2}{3}\int \frac{{\rm d}^3k}{(2\pi)^3}
\frac{k^2}{E_{i}^\star(k)}
(n_i(k)+\bar{n_i}(k)) -\frac{1}{2}m_\sigma^2\phi^2 \nonumber \\
-U(\phi) 
+\frac{g_{\omega}^2}{2m_\omega^2}\rho_B^2
+\frac{g_{\rho}^2}{2m_{\rho}^2}\rho_{B3}^2
-\frac{g_{\delta}^2}{2m_{\delta}^2}\rho_{S3}^2.
\end{eqnarray}

We remind that the baryon
densities $\rho_B$ are given by 
($\gamma$ is the spin/isospin multiplicity)

\begin{eqnarray}\label{eq.12}
\rho_B=\gamma
\int\frac{{\rm d}^3k}{(2\pi)^3}(n(k)-\bar{n}(k))\,,
\end{eqnarray}

while the scalar densities $\rho_S$ are

\begin{eqnarray}\label{eq.13}
\rho_S=\gamma
\int\frac{{\rm d}^3k}{(2\pi)^3}\frac{M^\star}{E^\star}(n(k)+\bar{n}(k)).
\end{eqnarray}

At the temperatures of interest here the antybaryon contributions
are actually negligible.

In order to study the asymmetric nuclear matter, we introduce
an asymmetry parameter $\alpha$ defined as 
$\alpha$ = $(\rho_{Bn}-\rho_{Bp})/\rho_B$ = $\frac{N-Z}{A}$.

The energy density and pressure for symmetric and asymmetric 
nuclear matter at finite temperature and the $n,p$ effective masses 
can be calculated self-consistently 
from Eqs. (6) to (13) just in terms of the four boson coupling constants,
$f_i \equiv (\frac{g_i^2}{m_i^2})$, $i = \sigma, \omega, \rho, \delta$
and the two parameters of the $\sigma$ self-interacting terms, 
$A \equiv \frac{a}{g_\sigma^3}$ and $B \equiv \frac{b}{g_\sigma^4}$. 

The isoscalar meson parameters are fixed from symmetric nuclear matter
properties at $T=0$: saturation density $\rho_0=0.16fm^{-3}$,
 binding energy $E/A = -16MeV$, nucleon effective mass $M^\star = 0.75 M_N$
 ($M_N=939MeV$) and incompressibility $K_V = 240 MeV$ at $\rho_0$. The fitted
$f_\sigma, f_\omega, A, B$ parameters are reported in Table I and have
quite standard values for these minimal non-linear $RMF$ models.

\vspace{1cm}
\noindent

{\bf Table I.} Parameter sets.

\par
\vspace{0.8cm}
\noindent
 
\begin{center}
\begin{tabular}{ c c c c } \hline
$parameter$   & $Set~I$         &~ $Set~II$   &~ $NL3$       \\ \hline
$f_\sigma~(fm^2)$  &10.33    &$same$        &~ 15.73  \\ \hline
$f_\omega~(fm^2)$  &5.42    &$same$         &~ 10.53  \\ \hline
$f_\rho~(fm^2)$    &0.95    &~3.15          &~ 1.34   \\ \hline
$f_\delta~(fm^2)$  &0.00     &~2.50         &~ 0.00   \\ \hline
$A~(fm^{-1})$     &0.033    &$same$         &~ -0.01  \\ \hline
$B$               &-0.0048  &$same$         &~ -0.003  \\ \hline
\end{tabular}
\end{center}

\vskip 1cm

In the table we report also the $NL3$ parametrization, widely used in
nuclear structure calculations \cite{lari97}. We remind that the
$NL3$-saturation properties for symmetric matter are chosen as 
$\rho_0 = 0.148fm^{-3}$, $M^\star = 0.6 M_N$, 
$K_V = 271.8 MeV$ and the symmetry parameter (see next section)
is $a_4 = 37.4 MeV$.

The corresponding phase diagram (pressure isotherms) is shown in Fig.1.
The local minimum disappears at the critical 
temperature ${\rm T}_{\rm c}$, determined by 
$\frac{\partial P}{\partial \rho}\vert_{{\rm T}_c}=\frac{\partial^2 P}{\partial
\rho^2}\vert_{{\rm T}_c}=0$. The obtained critical temperature of symmetric 
matter is ${\rm T}_{\rm c}=15.86$ MeV.

\begin{figure}[htb]
\epsfysize=6.5cm
\centerline{\epsfbox{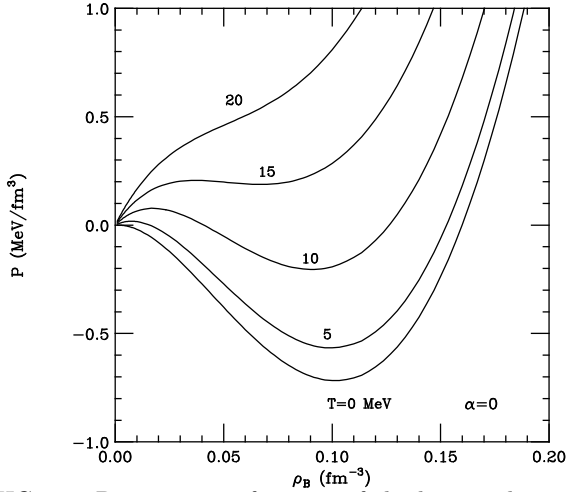}}
\caption{
Pressure as a function of the baryon density 
for symmetric nuclear matter ($\alpha$=0) at different temperatures.}
\end{figure}

\section{The symmetry energy}

Symmetry properties of asymmetric nuclear matter are univocally fixed from
the coupling constants of the isovector channels \cite{hartree}.
Experimentally we have just one relatively well known quantity, the
bulk symmetry energy $a_4$ of the Weiszaecker mass formula, in the
range $30-35~MeV$. In all the $RMF$ models with only the isovector $\rho$
meson this assigns the $f_\rho$ constant, that gives also the "slope"
of the symmetry term (the {\it Symmetry Pressure}) \cite{yst98}, apart small 
non-linear contributions \cite{mu96}.

When we further include the $\delta$-meson, from the $a_4$-value we will
fix only a combination of the two coupling constants, while the $\delta$
itself will imply interesting new contributions to the slope and
curvature of the symmetry energy and a neutron-proton effective mass
splitting. These will be the main points discussed in the following. We
note that most results are analytical and this will largely improve
the physical understanding of the effects due to the coupling to the 
$\delta$ isovector scalar channel.

The symmetry energy in asymmetric $NM$ is defined from the expansion
of the energy per nucleon $E(\rho_B,\alpha)$ in terms of the asymmetry
parameter
\begin{eqnarray}\label{eq.14}
E(\rho_B,\alpha)~\equiv~\frac{\epsilon(\rho_B,\alpha)}{\rho_B}~
=~E(\rho_B) + E_{sym}(\rho_B) \alpha^2 \nonumber \\
+ O(\alpha^4) +...
\end{eqnarray}
and so in general
\begin{equation}\label{eq.15}
E_{sym}~\equiv~\frac{1}{2} \frac{\partial^2E(\rho_B,\alpha)}
{\partial\alpha^2} \vert_{\alpha=0}~
=~\frac{1}{2} \rho_B \frac{\partial^2\epsilon}{\partial\rho_{B3}^2}
 \vert_{\rho_{B3}=0}
\end{equation}

\subsection*{The Bulk Symmetry Parameter}

The bulk symmetry parameter $a_4$ is just the value of $E_{sym}$
corresponding to the normal density $\rho_B=\rho_0$, at $T=0$.
 Then we discuss first the case at zero temperature, in order
to fix the isovector coupling constants.
>From Eqs.(10,15) we can easily get an
explicit expression for the symmetry energy \cite{ku97}
\begin{equation}\label{eq.16}
E_{sym}(\rho_B) = \frac{1}{6} \frac{k_F^2}{E_F} + \frac{1}{2}
f_\rho \rho_B - \frac{1}{2} f_\delta 
\frac{{M^\star}^2 \rho_B}{E_F^2 [1+f_\delta A(k_F,M^\star)]} 
\end{equation}
where $k_F$ is the nucleon Fermi momentum corresponding to $\rho_B$, 
$E_F \equiv \sqrt{(k_F^2+{M^\star}^2)}$ and $M^\star$ is the effective
nucleon mass in symmetric $NM$, $M^\star=M_N - g_\sigma \phi$.

The integral
\begin{equation}\label{eq.17}
A(k_F,M^\star)~\equiv~\frac{4}{(2\pi)^3} \int d^3k \frac{k^2}{ 
(k^2+{M^\star}^2)^{3/2}}
\end{equation}
has a simple analytical structure which makes quite transparent the
effect of the $\delta$-meson on the symmetry energy. With some algebra
we can get
\begin{equation}\label{eq.18}
A(k_F,M^\star) = 3 ( \frac{\rho_S}{M^\star} - \frac{\rho_B}{E_F})
\end{equation}
making use of
\begin{eqnarray}\label{eq.19}
\rho_S(T=0) = \frac{M^\star}{\pi^2} [k_FE_F - {M^\star}^2 \ln
(\frac{k_F+E_F}{M^\star})] \nonumber \\
\rho_B(T=0) = \frac{2}{{3\pi}^2} k_F^3
\end{eqnarray}
Expanding the scalar density in terms of $(k_F/M^*)^2$  \cite{sewa86}, the
quantity Eq.(18) can be written as
\begin{eqnarray}
A(k_F,M^\star) = \frac{3 \rho_B}{M^*}
[\frac{1}{5}(\frac{k_F}{M^*})^2 - 
 \frac{3}{14}(\frac{k_F}{M^*})^4 
- \frac{5}{24}(\frac{k_F}{M^*})^6 + ...], \nonumber
\end{eqnarray}
that can be used to derive a similar expansion for the symmetry energy
eq.(16).

We see that $A(k_F,M^\star)$ is certainly very small at low densities, 
below $\rho_0$.
It can be still neglected up to a baryon density 
$\rho_B \simeq 3\rho_0$, where it reaches the value $0.045 fm^{-2}$
(with our symmetric $NM$ parameters), i.e. a correction of about $10\%$
in the denominator of Eq.(16).

Then in the density range of interest here we can use, at the leading order,
the much simpler 
form of the symmetry energy, extremely
nice in order to get the main qualitative new features due to 
the $\delta$-meson
coupling:
\begin{equation}\label{eq.20}
E_{sym}(\rho_B) = \frac{1}{6} \frac{k_F^2}{E_F} + \frac{1}{2}
[f_\rho - f_\delta 
(\frac{M^\star}{E_F})^2] \rho_B
\end{equation}
We see that, when the $\delta$ is included, the observed $a_4$ value actually
assigns the combination $[f_\rho - f_\delta (\frac{M^\star}{E_F})^2]$
of the $(\rho,\delta)$ coupling constants. If
$f_\delta \not= 0$ we have to increase the $\rho$-coupling (see Fig.1 of
\cite{ku97}). In our calculations we use the value $a_4 = 30.5 MeV$.
In Table I the Set I corresponds to $f_\delta=0$. In the Set II
$f_\delta$ is chosen as $2.5fm^2$. Although this value is relatively
well justified, \cite{bonn}, we stress that aim of this work is just
to show the main qualitative new effects of the $\delta$-coupling.

In order to have the same $a_4$ we must increase the $\rho$-coupling constant
of a factor three, up to $f_\rho=3.15fm^2$. From the structure of 
Eqs.(16, 20) it is clear that there is a connection between the scalar field
(isoscalar, $\sigma$, and isovector, $\delta$) contributions, since
both are acting on effective masses. A strong $\sigma$-coupling 
(smaller $M^*$, e.g. see NL3 parametrization) can compensate
a strong $\delta$-coupling. In our evaluation we keep fixed the $\sigma$ 
parameters ($f_\sigma, A, B$) leading to $M^*=0.75M_N$. In any case 
a larger value of $E_{sym}$ at high baryon densities, due to 
the relativistic
mechanism discussed in the following, see Fig.2, will be always present
when the $\delta$ is included.

At subnuclear densities, $\rho_B < \rho_0$, in both cases, 
$(\rho)$ and $(\rho+\delta)$, from Eq.(20) we have an almost 
linear dependence 
of $E_{sym}$ on
the baryon density, since $M^\star \simeq E_F$ as a  good approximation .
Around and above $\rho_0$ we expect a steeper increase in the
$(\rho+\delta)$ case since $M^\star/E_F$ is decreasing, see Fig.2.

\begin{figure}[htb]
\epsfysize=6.5cm
\centerline{\epsfbox{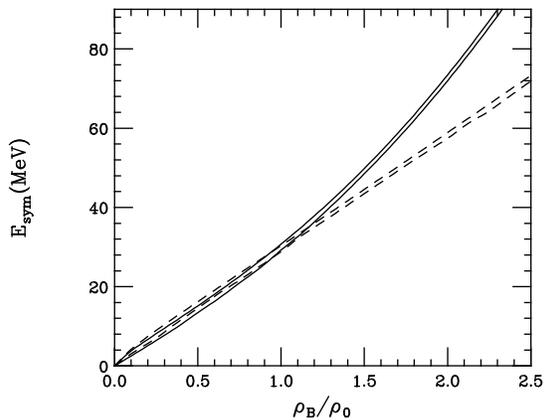}}
\caption{
Total ($kinetic~+~potential$) symmetry energy as a 
function of the baryon density.
 Dashed lines: $(\rho)$. Solid lines: $(\rho+\delta)$.
 Upper curves: zero temperature. Lower curves: $T=8MeV$.}
\end{figure}

This is 
an interesting aspect to look at more in detail since it actually represents
a general relativistic effect due to the coupling to scalar mesons.
In some sense it is the equivalent, in the isovector channel ($\delta~vs.~\rho$),
i.e. for the symmetry energy, of the saturation mechanism we have in the 
isoscalar channel ($\sigma~vs.~\omega$)
for the symmetric matter.

The scalar charged $\delta$-meson, like the neutral $\sigma$, is acting
on the nucleon effective masses, introducing a $n,p$ splitting, see
Eq.(6). This causes a negative contribution to $E_{sym}$, Eqs.(16),(21),
since it reduces the gap between $n,p$ Fermi energies due to the
different Fermi momenta in asymmetric $NM$. In fact in a n-rich matter
the neutron Fermi momentum increases while the neutron effective mass
decreases, see Eq.(6) and Fig.6 (the opposite for protons).

Such negative $\delta$ contribution is reduced at high densities
due to the "Lorentz contraction" factor $(M^\star/E_F)^2$,
 Eqs.(16) and (20), that in general gives the attenuation of the
scalar interactions with increasing baryon density.

We stress the consistent picture of a symmetry energy
built from the
balance of scalar (attractive) and vector (repulsive) contributions,
with the scalar channel becoming weaker with increasing baryon density.
This is indeed the isovector counterpart of the saturation mechanism 
occurring in the isoscalar channel for the symmetric nuclear matter.
 From such
scheme we get a further strong fundamental support for the introduction
of the $\delta$-coupling in the symmetry energy evaluation.

\begin{figure}[htb]
\epsfysize=6.7cm
\centerline{\epsfbox{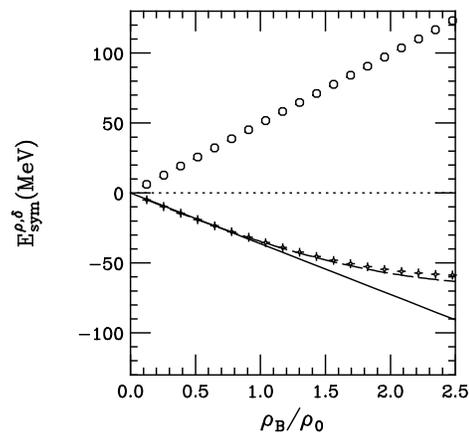}}
\caption{
$\rho$- (open circles) and $\delta$- (crosses) contributions
to the symmetry energy, second and third terms of Eq.(16). The dashed line 
is the approximate $\delta$-contribution of Eq.(20), see text.
The solid line is the linear extrapolation of the 
low density behaviour, plotted to guide the eye.}
\end{figure}

In Fig.3, using our parametrizations, we quantitatively show the
interplay of the two contributions ($\rho~and~\delta$) to
$E_{sym}$, second and third terms of Eqs.(16),(20), as a function
of the baryon density. The crosses (negative $\delta$ contribution) follow
a linear behaviour up to roughly $\rho_0$ and then they tend to saturate
due to the Lorentz contraction factor. In correspondence we will see the 
increase of the total $E_{sym}$ shown in Fig.2. We note the accuracy of the
approximate $\delta$ contribution given by Eq.(20), dashed line in Fig.3.

In Fig.4 we show the Equation of State (energy per nucleon) for pure
neutron matter ($\alpha=1$) obtained with the two parameter Sets, 
I ($\rho$) and II ($\rho+\delta$).

The values are in good agreement with recent non-relativistic 
Quantum-Monte-Carlo
variational calculations with realistic $2-$ and $3-$body
forces \cite{fan01}. The inclusion of the $\delta$-coupling
leads to a larger repulsion at baryon densities roughly above $1.5\rho_0$.
This could be of interest for the structure of neutron stars
and the possibility of a transition to new forms of nuclear matter.

\begin{figure}[htb]
\epsfysize=4.0cm
\centerline{\epsfbox{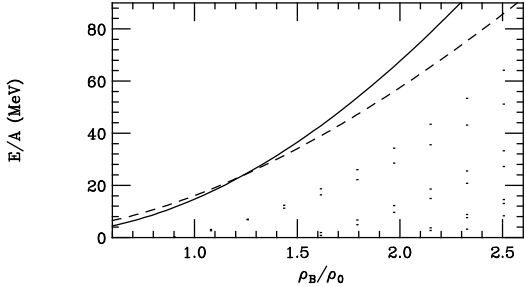}}
\caption{
$EOS$ for pure neutron matter.
 Dashed line: $(\rho)$. Solid line: $(\rho+\delta)$.}
\end{figure}

>From the previous analysis we expect to see also 
interesting $\delta$-effects on
the slope ({\it symmetry pressure}) and curvature ({\it symmetry 
incompressibility})
of the symmetry energy around $\rho_0$, of relevant physical meaning.
This will be the main subject of the following discussion.

\subsection*{Symmetry Pressure and Symmetry Incompressibility}

In order to have a quantitative evaluation of the effect on the slope
and curvature, and also to compare with other estimates, we use
the expansion of the symmetry energy around $\rho_0$ \cite{loqu88,baocr}
\begin{equation}\label{eq.21}
E_{sym}(\rho_B) = a_4 + \frac{L}{3} (\frac{\rho_B-\rho_0}{\rho_0})
 + \frac{K_{sym}}{18} (\frac{\rho_B-\rho_0}{\rho_0})^2 + ...
\end{equation} 
with $L$ and $K_{sym}$ respectively related to slope and curvature of 
the symmetry energy at $\rho_0$
\begin{eqnarray}\label{eq.22}
L = 3\rho_0 (\frac{\partial E_{sym}}{\partial\rho_B}) \mid _{\rho_B=\rho_0}
~~~and~~~ \nonumber \\
K_{sym} = 9\rho_0^2 (\frac{\partial^2 E_{sym}}{\partial\rho_B^2})
 \mid _{\rho_B=\rho_0}
\end{eqnarray}
>From Eq.(20) we get a potential contribution to the density variation
of $E_{sym}$ given by (after some algebra)
\begin{eqnarray}\label{eq.23}
\frac{\partial E_{sym}}{\partial\rho_B} \mid _{pot} =
 \frac{1}{2} [f_\rho - f_\delta (\frac{M^\star}{E_F})^2] +
 \nonumber \\ 
f_\delta (\frac{M^\star k_F}{E_F^2})^2 [\frac{1}{3} - 
 \frac{\rho_B}{M^\star}\frac{\partial M^\star}{\partial\rho_B}]
\end{eqnarray}
Around normal density $\rho_0$ the first term is fixed by the
$a_4$ value and so we get a net increase of the slope, due to the $\delta$,
given by the second term (always positive since
$({\partial M^\star}/{\partial\rho_B}) < 0$).

>From the first term of Eq.(23), with our parametrization, we get a potential
contribution to $L$ of $45MeV$ (this is actually fixed by the $a_4$ value)
and a genuine $\delta$ extra contribution of about $20MeV$ from the 
second term. When we include also the kinetic part (from the
first term of Eq.(20)) we have a total slope parameter going from
$L(\rho)=+84MeV$ to $L(\rho+\delta)=+103MeV$. 

We note again that the slope parameter, or equivalently the
{\it Symmetry Pressure} $P_{sym} \equiv \rho_0 L/3$, is of great
importance for structure properties, being linked to the
thickness of the neutron skin in n-rich (stable and/or unstable)
nuclei \cite{pera95,bro00,typ01,hor01}, and to the assessment
of the drip-line \cite{legm01}. Moreover the same parameter
gives an estimate of the shift of the saturation density
with asymmetry (at the lowest order in $\alpha^2$)
\begin{equation}\label{eq.24}
\Delta \rho_0 (\alpha) = - \frac{3\rho_0L}{K_V(\alpha=0)} \alpha^2
\end{equation}
that can be easily obtained from a linear expansion around
the symmetric value $\rho_0(\alpha=0)$ \cite{eri98}. 

Eq.(24) has a simple
physical meaning: in order to compensate the symmetry pressure
in asymmetric matter we have to move  the zero 
of the $P(\rho_B)$
curve to lower densities.
The amount of the shift will be inversely proportional to the slope,
given by the incompressibility.

\begin{figure}[htb]
\epsfysize=6.0cm
\centerline{\epsfbox{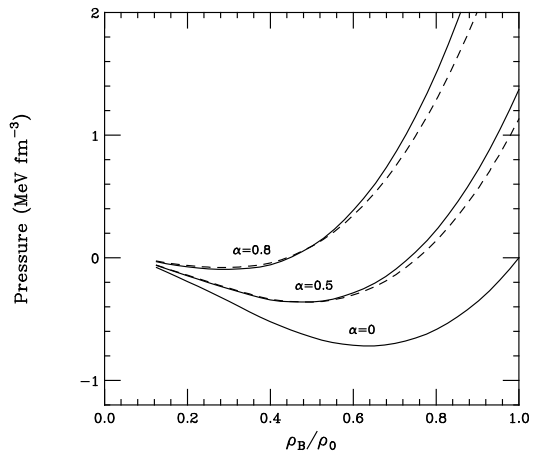}}
\caption{
Pressure as a function of the baryon density at zero
temperature
for various asymmetries $\alpha=0~(bottom), 0.5~(middle), 0.8~(top)$.
 Dashed lines: $(\rho)$. Solid lines: $(\rho+\delta)$.}
\end{figure}

All that can be seen in Fig.5  where we report
the total pressure $P(\rho_B)$ for various asymmetries,
at zero temperature, in the two cases, $\rho$-meson only
(dashed lines) and $(\rho+\delta)$ (solid lines).

It is instructive to perform a similar discussion for the
curvature parameter $K_{sym}$, Eq.(21). Now the potential
contribution is {\it exclusively} given by the $\delta$-meson,
with a definite positive sign, as we can see from the previous 
discussion.
This will be a large effect on the total since the kinetic
part is quite small \cite{curvature}. 
If we compare with
non-relativistic effective parametrizations \cite{bkb98,baocr},
when we add the $\delta$-meson
we move from a linear to a roughly parabolic $\rho_B$-dependence of
the symmetry energy.

With our parameters we pass from a $K_{sym}(\rho) = +7 MeV$
(only kinetic) to a $K_{sym}(\rho+\delta) = +120 MeV$.
So this quantity appears extremely interesting to look at experimentally,
as recently suggested from reaction measurements \cite{baocr}.
The problem is that the effect on the total incompressibility
of asymmetric matter, that likely could be easier to measure,
is not trivial. We can evaluate a shift of the incompressibility
with asymmetry, at the same $\alpha^2$ order as in Eq.(24), as
\cite{loqu88,eri98}
\begin{equation}\label{eq.25}
\Delta K_V(\alpha) = (K_{sym}-6L) \alpha^2
\end{equation} 
So what really matters for the total incompressibility is the 
combination $(K_{sym}-6L)$, with the possibility of a compensation 
between the two terms. Just by chance this is actually what is
happening in
our calculations since for the above combination we get
$-497 MeV$ in the case of only $\rho$ coupling, and $-504 MeV$
when we add also the $\delta$. Indeed we see from Fig.5 that
the slopes of the pressure around equilibrium density 
($P(\rho_0)=0$) are very close
in the two cases, $(\rho)$ and $(\rho+\delta)$.

\subsection*{Effective Mass Splitting}

Another qualitative new result of the $\delta$-meson coupling is
the $n,p$-effective mass splitting in asymmetric matter \cite{ku97},
see Eq.(6). In Fig.6 we report the baryon density dependence of the
$n,p$ effective masses for $\alpha=0.5,~(N=3Z)$ asymmetry, calculated 
with our Set II parameters, compared to the symmetric case \cite{rhos}.

\begin{figure}[htb]
\epsfysize=6.7cm
\centerline{\epsfbox{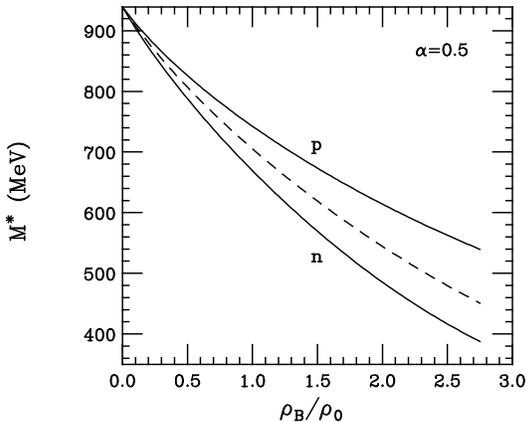}}
\caption{
Neutron, proton effective masses vs. the baryon density 
for $\alpha=0.5$ ($N=3Z$). The dashed line corresponds to symmetric nuclear
matter.}
\end{figure}

We get a splitting of the order of $15\%$ at normal density $\rho_0$,
and increasing with baryon density. Unfortunately from the present nuclear 
data we have a very small knowledge of this effect, due to the low
asymmetries available. This issue will be quite relevant in the
study of drip-line nuclei. Moreover we can expect important effects
on transport properties ( fast particle emission, collective flows)
 of the dense and asymmetric $NM$ that will
be reached in Radioactive Beam collisions at intermediate energies.

The sign itself of the splitting would be very instructive. As we
can see from Eq.(6) in $n$-rich systems we expect a neutron effective
mass always smaller than the proton one. The same is predicted
from more microscopic relativistic Dirac-Brueckner calculations
\cite{hole01}. At variance, non-relativistic Brueckner-Hatree-Fock
calculations are leading to opposite conclusions \cite{bom01,zbl99}.
Still in the non-relativistic picture quite contradictory results
are obtained with the Skyrme effective forces.
The most recent 
parametrizations, SLy-type 
\cite{Cha97}, of Skyrme forces give the proton effective mass
above the neutron one, in agreement with our calculations.
Previous Skyrme-like forces, instead, yield a splitting in the
opposite direction, but also show unpleasant behaviours in the spin channel
(collapse of polarized neutron matter, see discussion in \cite{Cha97}). 

We like to note again that, as shown before, a decreasing neutron 
effective mass in $n$-rich matter is behind the relativistic mechanism
for the symmetry energy, balance of scalar (attractive) and vector (repulsive)
contributions in the isovector channel.

\subsection*{Finite Temperature Effects}

In the temperature range of interest in this paper, below the
critical temperature $T_c$ of the liquid-gas phase transition,
of the order of $15-16$ MeV (see Fig.1), temperature effects
on the symmetry properties are not expected to be large. Indeed
the contributions of antifermions, that could modify all the terms
with the scalar densities, see Eq.(13), are still very reduced.
In Fig.2 we present also the symmetry energies calculated at
$T=8MeV$, lower solid and dashed curves. For both cases, $\rho$ and 
$(\rho+\delta)$ the variation is quite small. We have a reduction
mainly coming from the kinetic contribution due to the smoothing of the
$n,p$ Fermi distributions. We note that this result is in full agreement
with relativistic Brueckner-Hartree-Fock calculations \cite{hww97}. 

The effect is even smaller for the mass splitting, given by a difference 
of scalar densities. In the next section we will study in detail the
phase diagram of heated asymmetric nuclear matter, focussing in
particular on the instability regions.

\section{Mechanical and chemical instabilities}

Heavy-ion collisions can provide the possibility of studying
equilibrated nuclear matter far away from normal conditions,
i.e. to sample new regions of the $NM$ phase diagram.
In particular the process of multifragmentation allows to probe
dilute nuclear matter at finite temperatures, where in the symmetric case,
see Fig.1, we expect to see
a phase transition of first order of liquid-gas type, as suggested
from the very first equations of state built with effective interactions
\cite{mos76,ber83,mek84}.

The multifragmentation phenomenon may be interpreted as a signal of
such a phase transition in a finite system, when the nuclear matter 
in the expansion phase enters the region of spinodal (mechanical)
instability.
Nuclear matter is however a two-component system consisting of
neutrons and protons: binary systems have more complicated phase diagrams
due to the new concentration degree of freedom. In particular
for asymmetric nuclear matter
a qualitative new feature in the liquid-gas phase transition
is expected, the onset of a coupling to chemical instabilities 
(component separation) that will show
up in a novel nature of the unstable modes, the mixture of
density and charge fluctuations leading to an Isospin Distillation
effect \cite{mu95,bao197,bar01}. Indeed equilibrium thermodynamics, 
as well as
non-equilibrium kinetic considerations predict
that an asymmetric system will separate into more symmetric larger
fragments ("liquid") and into neutron-rich light fragments ("gas").
So the chemical instability can be investigated experimentally
just by measuring the N/Z ratio, or isospin content, of the
fragments. 

Since the effect is driven by the isospin dependent part of the
nuclear equation of state, here we will look at the influence of
the $\delta$ coupling on this new liquid-gas phase transition.

It is known that the stability condition
 of a two-component, $n-p$, thermodynamical system is given by

\begin{eqnarray}\label{eq.26}
\left(\frac{\partial P}{\partial \rho}\right)_{T,y}
\left(\frac{\partial \mu_{p}}{\partial y}\right)_{T,P} > 0,
\end{eqnarray}
where P is the pressure, $\mu_{p}$ is the proton chemical potential 
and $y$ the proton fraction $Z/A$, related to the asymmetry parameter
$\alpha=1-2y$.

Eq.(26) is equivalent to set the free energy
to be a convex function in the  space
of the $n,p$ density oscillations, $\delta\rho_n,\delta\rho_p$.
In charge symmetric matter
{\it isoscalar} (total density) $\delta\rho_n+\delta\rho_p$ and 
{\it isovector} (concentration)
 $\delta\rho_n-\delta\rho_p$ oscillations are not coupled and we
have two separate conditions for instability
\begin{eqnarray}\label{eq.27}
\left(\frac{\partial P}{\partial \rho}\right)_{T,y} \leq 0, 
\end{eqnarray}
{\it mechanical}, i.e. vs. density oscillations (the spinodal region), and

\begin{eqnarray}\label{eq.28}
\left(\frac{\partial \mu_{p}}{\partial y}\right)_{T,P} \leq 0,
\end{eqnarray}
{\it chemical}, i.e. vs. concentration oscillations.

In asymmetric matter the isoscalar and isovector modes are coupled
and the two separate inequalities do not maintain anymore a physical 
meaning, in the sense that they not select the nature of the instability.
Inside the general condition Eq.(26) the corresponding
unstable modes are a mixing of density and concentration oscillations,
very sensitive to the charge dependent part of the nuclear interaction
in the various instability regions \cite{bar01}.

In dilute asymmetric $NM$ (n-rich) the normal unstable modes for all
realistic effective interactions are still {\it isoscalarlike},
i.e. in phase $n-p$ oscillations but with a larger proton
component \cite{bar01}. This leads to a more symmetric high density
(liquid) phase everywhere under the instability line defined by
Eq.(26) and consequentely to a more neutron-rich gas ({\it Isospin 
Distillation}). Such "chemical effect" is driven by the increasing 
symmetry repulsion going from low density to roughly the saturation
value and so it appears quite
sensitive to the symmetry energy of the used effective interaction 
at subnuclear densities. It could provide a good opportunity to
differentiate the various $EOS$ isospin dependences \cite{isov}.

In this section we study  the effect of the $\delta$-coupling on
the instability region given by Eq.(26) in dilute asymmetric nuclear
matter  and on the structure of the corresponding unstable modes.
We start from an identity valid for any binary thermodynamical system
\begin{eqnarray}\label{eq.29}
\left(\frac{\partial \mu_{p}}{\partial \rho_{p}}\right)_{T,\rho_{n}} 
\left(\frac{\partial \mu_{n}}{\partial \rho_{n}}\right)_{T,\rho_{p}}-
\left(\frac{\partial \mu_{p}}{\partial \rho_{n}}\right)_{T,\rho_{p}} 
\left(\frac{\partial \mu_{n}}{\partial \rho_{p}}\right)_{T,\rho_{n}} =
 \nonumber \\
\frac{1}{(1-y)\rho^2}
\left(\frac{\partial P}{\partial \rho}\right)_{T,y}
\left(\frac{\partial \mu_{p}}{\partial y}\right)_{T,P} 
\end{eqnarray}
where $\mu_{q}$,  $\rho_{q}$ ($q = n,p$) are 
respectively neutron/proton
 chemical potentials and densities. From our knowledge of the
chemical potential on each isotherm, Eq.(9), we can easily 
compute the limits of the instability region in the $T,\rho_B$
plane for dilute asymmetric $NM$ in the two choices, without and with
the $\delta$ meson.

The results are reported in Fig.7. The $\delta$ inclusion (solid lines)
appears not much affecting the instability limits even at relatively
large asymmetry $\alpha=0.8$ ($N \simeq 9Z$). We notice just a 
small reduction and a shift
to the left (lower densities) of the whole region: this can be understood
in terms of the larger symmetry repulsion, see Fig.3.

\begin{figure}[htb]
\epsfysize=6.7cm
\centerline{\epsfbox{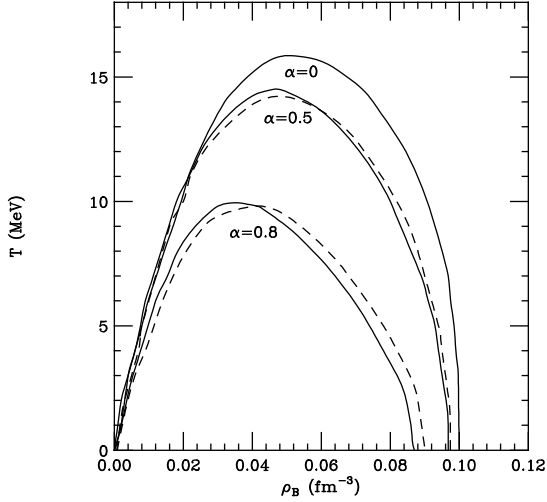}}
\caption{
Limits of the instability region in the $T,\rho_B$ plane
for various asymmetries.
Dashed lines: $(\rho)$. Solid lines: $(\rho+\delta)$ [for $\alpha \not= 0$].}
\end{figure}

We can understand the relatively small $\delta$ effect on the stability 
border just remembering that for low densities, well below $\rho_0$, 
the symmetry term has roughly the same linear behaviour in both $(\rho)$
and $(\rho+\delta)$ schemes, fixed by the $a_4$ parameter (see the
discussion after Eq.(20)).  
Only for very large asymmetries
it appears
relatively easier in the $\delta$ case to be 
in the stable liquid 
phase.

A rather larger difference can be seen  in the behaviour of the
quantity Eq.(26) inside the instability region.
This is plotted in Fig.8 for various asymmetries at zero temperature
and in Fig.9 at various temperatures for a fixed asymmetry $\alpha=0.5$
($N=3Z$). The solid curves (with $\delta$-coupling) are systematically
above the dashed ones, signature of a weaker instability.

\begin{figure}[htb]
\epsfysize=6.7cm
\centerline{\epsfbox{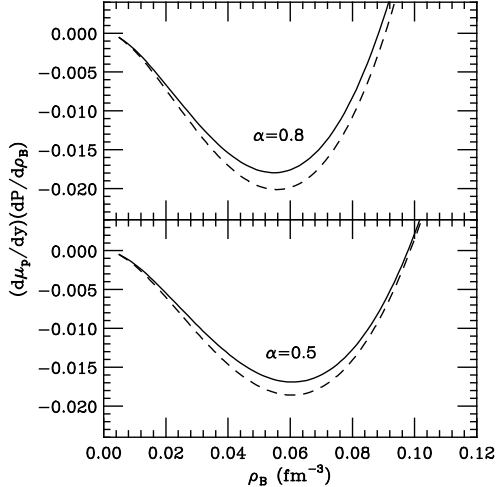}}
\caption{
The quantity Eq.(26) inside the instability region
at $T=0$ and various asymmetries.
Dashed lines: $(\rho)$. Solid lines: $(\rho+\delta)$.}
\end{figure}

\begin{figure}[htb]
\epsfysize=6.2cm
\centerline{\epsfbox{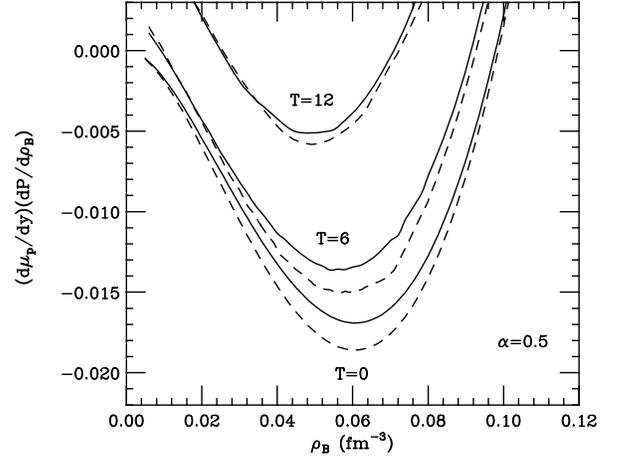}}
\caption{
The same as in Fig.8 at various temperatures and
fixed asymmetry $\alpha=0.5$.
Dashed lines: $(\rho)$. Solid lines: $(\rho+\delta)$.}
\end{figure}

In order to better understand the origin of such effect we study
also the structure of the corresponding unstable modes. We follow the 
Landau dispersion relation approach to small amplitude oscillations
in Fermi liquids \cite{landau,bay91,cdl98,bar98}. For a two component
($n,p$) matter the interaction is characterized by the Landau
parameters $F_0^{q,q'}, (q,q')=(n,p)$ defined as
\begin{equation}\label{eq.30}
N_q(T) \frac{\partial\mu_q}{\partial\rho_{q'}} \equiv \delta_{q,q'}
 + F_0^{q,q'}
\end{equation} 
where $N_q(T)$ represents the single particle level density
 at the Fermi energy. At zero temperature it has the simple form
$$
N_q = \frac{k_{Fq} E^*_{Fq}}{\pi^2},~~~~~q = n, p.
$$

In the symmetric case ($F_0^{nn}=F_0^{pp}, F_0^{np}=F_0^{pn}$), the 
Eqs.(27,28) correspond to the two Pomeranchuk instability conditions
\begin{eqnarray}\label{eq.31}
F_0^s = F_0^{nn} + F_0^{np} < -1 ~~~~mechanical \nonumber \\
F_0^a = F_0^{nn} - F_0^{np} < -1 ~~~~chemical.
\end{eqnarray} 
>From the dispersion relations $F_0^s$ will give the properties of density
(isoscalar) modes and $F_0^a$ of the concentration (isovector) modes.
For asymmetric $NM$ we have some corresponding generalized Landau
parameters $F_{0g}^s, F_{0g}^a$ which will characterize the new collective
response (respectively {\it isoscalarlike} and {\it isovectorlike}). They
can be expressed as 
fixed combinations of the $F_0^{q,q'}$ for each baryon density,
asymmetry and temperature. This transformation
is reducing Eq.(29) to a "diagonal" form  \cite{bar01}
\begin{eqnarray}\label{eq.32}
(1+F_{0g}^s)(1+F_{0g}^a) = 
\frac{4}{(1-y)\rho^2}\left(\frac{N_nN_p}{N_n+N_p}\right)^2
 \nonumber \\
\times\left(\frac{\partial P}{\partial \rho}\right)_{T,y}
\left(\frac{\partial \mu_{p}}{\partial y}\right)_{T,P} 
\end{eqnarray}

As already discussed in the unstable region of dilute asymmetric $NM$
we have {\it isoscalarlike} unstable modes and so $1+F_{0g}^s<0$,
while the combination $1+F_{0g}^a$ will stay positive.
In Fig.10 we report the full calculation of these two quantities
in the unstable region at zero temperature, for asymmetry $\alpha=0.5$,
 with and without
the $\delta$ coupling.

\begin{figure}[htb]
\epsfysize=6.5cm
\centerline{\epsfbox{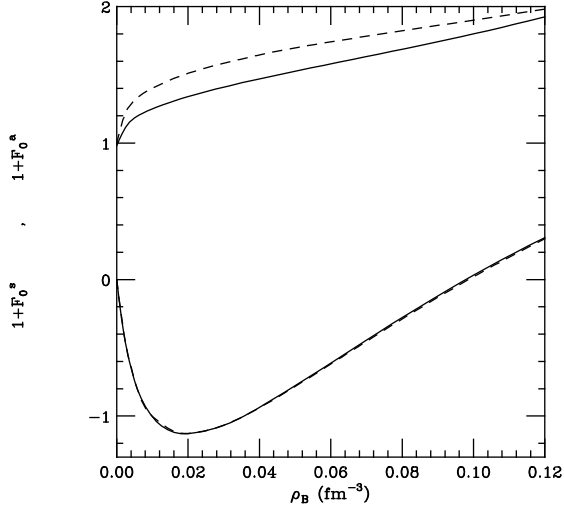}}
\caption{
Behaviour of the generalized Landau parameters
inside the instability region at zero temperature and
asymmetry $\alpha=0.5$ ($N=3Z$).
Dashed lines: $(\rho)$. Solid lines: $(\rho+\delta)$.}
\end{figure}

The $\delta$ meson is almost not affecting at all the unstable mode,
given by the $F_{0g}^s$ parameter, and so the limits of the instability
region, for $\alpha=0.5$, are not changed (see the Figs.7,8,9). We have a 
much larger
effect on the $F_{0g}^a$ parameter which describe
"stable" {\it isovectorlike} modes, that actually can propagate
as good zero-sound collective motions since $F_{0g}^a>0$.
This can be expected from the isovector nature of the $\delta$-meson. 
>From here
we get the main differences seen in Figs.8,9 for the quantity Eq.(26) 
inside the
instability region, just using Eq.(32).

Another interesting aspect of the comparison between Fig.8 and Fig.10
is the shift of the "maximum instability" density region. From the
thermodynamical condition reported in Fig.8 it seems that the
largest instability (the most negative value) is around 
$\rho_B=0.06fm^{-3}$. In fact from Fig.10 we see that the fastest
unstable mode, corresponding to the most negative Pomeranchuk
condition for $1+F_{0g}^s$, is actually present for more
dilute matter, around $\rho_B=0.02fm^{-3}$. This shows the
importance of  the linear response analysis.

Finally the fact that the $\delta$-coupling is mostly affecting
the {\it isovectorlike} modes is of great interest for
possible effects on the isovector Giant Dipole Resonances
studied around normal density within the $RMF$ approach
in asymmetric systems. 

\section{Conclusion and outlook}

Aim of this work has been to select genuine new effects 
on properties of the asymmetric nuclear matter due to the coupling to the
charged $\delta$-scalar meson introduced in a minimal way
in a phenomenological hadronic field theory. 
Simple analytical results allow to clarify the interplay between
the $\rho$ and $\delta$ isovector meson contributions.

With respect to the Equation of State, when the $\delta$ is included
the symmetry energy keeps an almost linear baryon density repulsive behaviour
at sub-nuclear densities while it starts to move to a roughly parabolic
trend above the saturation density $\rho_0$. Such effect comes directly from
the general relativistic property of the weakening of the attractive
scalar interactions at high density. It represents the equivalent in
the isovector channel (interplay of $\rho$ and $\delta$ contributions)
of the saturation mechanism of the symmetric $NM$
(interplay of $\omega$ and $\sigma$ contributions).
All that gives a further fundamental support to the introduction
of the $\delta$-channel in the symmetry energy evaluation.

Such "$\delta$ mechanism" for the symmetry energy leads to a more
repulsive $EOS$ for pure neutron matter at baryon densities
roughly above $2\rho_0$. This is the region where transitions to
different forms of nuclear matter are expected and so the
result appears quite stimulating.

We have shown in details that the
new $\delta$ contributions are not negligible for the slope
parameter around $\rho_0$ (the {\it symmetry pressure}) and
absolutely essential for the curvature parameter 
({symmetry incompressibility}). The proton-neutron effective mass splitting
is also directly given by the $\delta$ coupling and appears to be
of the order of $15\%$ at normal nuclear density (for a $N=3Z$ asymmetry).

The possibility of an experimental observation of such effects is
suggested. We list here some sensitive observables:
\begin{itemize}
\item{Neutron distributions in $n-$rich nuclei (stable and unstable),
\cite{bro00,typ01,hor01}}
\item{Assessment of drip-line stabilities \cite{hole01,legm01}}
\item{Bulk densities and incompressibility modulus in asymmetric nuclei
 \cite{eri98,loqu88,baocr}}
\item{Transport properties in Radioactive Beam collisions at
intermediate energies: dissipative mechanisms, fast nucleon emission,
collective flows \cite{nova01,fab98,bao00,ditba,bkb98,scd99}}
\end{itemize} 

Effects on the new critical properties of warm $ANM$, mixing of mechanical 
and chemical instabilities and isospin distillation, are also presented.
The border of the instability regions as well as the nature of
the unstable fluctuations in dilute asymmetric matter
are not much affected. Indeed in the low density region the
symmetry energy has a very similar behaviour with and without
the $\delta$ meson.
This reduces the possibility of the observation of
$\delta$-coupling effects from measurements of isospin effects
in multifragmentation events in asymmetric heavy ion collisions.
We note however that other fragment production mechanisms,
like {\it "neck fragmentation"}, are expected to be very sensitive to
the symmetry potential around normal density  
since clusters are now formed in regions in contact
with the "spectator" matter \cite{ditcr,ditba}. We can predict some
interesting
effects on this kind of events, that actually represent a quite large part
of the fragment production cross section. 
 
>From the study of the $\delta$ influence on the Landau parameters
in the low baryon density region we note however an interesting
effect on the collective response of asymmetric $NM$. While the
unstable ({\it isoscalarlike}) modes are almost not modified,
the {\it isovectorlike} ones, which appear to have a good stable
zero-sound propagation, are quite sensitive to the introduction 
of the $\delta$. This suggests some importance of the $\delta$
meson in the $RMF$ description of isovector giant resonances
in asymmetric nuclei. 

In conclusion we would like to add a comment on the limits 
of the Hartree scheme used in this work, as well as in most
$RMF$ calculations. Indeed it is well known that exchange terms
can give contributions in the isovector scalar channel 
also in absence of explicit $\delta$-couplings, see ref.\cite{gre01}
and ref.s therein. The comparison with Dirac-Brueckner-Hartree-Fock
results of ref.\cite{hole01} and some Non-Linear-Hartree-Fock calculations
performed in \cite{gre01}, \cite{grepri}, are showing that
an effective $\delta$ coupling parameter $f_\delta \simeq 2.5fm^2$,
almost constant in a wide density range, is well accounting for the Fock 
term contributions. This makes us more confident on the reliability
of the quantitative evaluation of the $\delta$ effects discussed 
in this paper.

\vskip 3.0cm

\subsection*{Acknowledgments}

One of authors (L.B.) would like to thank the LNS in Catania 
for the hospitality during his stay.
This work was supported by the INFN of Italy and the National Natural
Science Foundation of China.


\begin{thebibliography}{99}
\bibitem{self} We like to remark that in general non-linear terms in 
the isoscalar vector channel $\omega$
appears to be less important, see \cite{mu95,fuse00}, as also confirmed
the good nuclear structure results of the $NL3$ parametrization \cite{lari97}.
Also in the case of the isovector vector channel $\rho$ non-linear terms
have been shown to have small effects \cite{mu96}.
\bibitem{hartree} This is the case of a $RMF$ model in the Hartree
approximation, mostly followed in all applications. For a discussion
on the effects of the Fock terms see the recent work \cite{gre01}
and ref.s therein.
\bibitem{bonn} This choice corresponds to the $\delta$ parameters
 ($g_\delta,m_\delta$) of the $Bonn~C$ potential \cite{mac89}.
We note that such value is also in quite good agreement with the
density dependent study of \cite{hole01} derived from a microscopic
Dirac-Brueckner-Hartree-Fock calculation in asymmetric matter 
(the plateau in Fig.2
of that ref.). In any case we are interested in the new qualitative features 
of the $\delta$ coupling.
\bibitem{curvature} We note that in the non-relativistic case the
kinetic contribution to $K_{sym}$ is actually negative, since it has
the form $-\pi^2\rho_0/(Mk_F)$ which is of the order $\sim -30 MeV$. 
In the relativistic 
case it becomes slightly positive due to the density dependence of the
nucleon effective mass.
\bibitem{rhos} The solid lines in Fig.4 are not fully symmetric with
respect to the $\alpha=0$ dashed line. This is due to the fact that,
for fixed baryon density, the total scalar density which gives
the $\sigma$ field contribution (first part of the Eq.(6)) is
also slightly $N/Z$ dependent.
\end{thebibliography}
\end{document}